\documentclass[nofootinbib,prd, preprint]{revtex4-1}
\usepackage{graphicx}
\usepackage{longtable}
\usepackage{amsmath}
\usepackage{amssymb}
\usepackage{subfigure}
\usepackage{hyperref}
\usepackage{stackengine}
\usepackage{color}

\newcommand{\dd}{\, {\rm d}}
\newcommand{\gsim}{\;\mbox{\raisebox{-0.5ex}{$\stackrel{>}{\scriptstyle{\sim}}$}
}\;}

\newcommand{\ve}{\varepsilon}

\newcommand{\gbt}{\hat{g}_3}
\newcommand{\gbf}{\hat{g}_4}

\newcommand{\so}{{\Phi_1}}
\newcommand{\rs}{\rho_s}

\DeclareMathOperator{\sech}{sech}

\begin{document}
\title{Oscillons in Higher-Derivative Effective Field Theories}
\author{Jeremy Sakstein}
\email[Email: ]{sakstein@physics.upenn.edu}
\affiliation{Center for Particle Cosmology, Department of Physics and Astronomy, University of Pennsylvania 209 S. 33rd St., Philadelphia, PA 19104, USA}
\author{Mark Trodden}
\email[Email: ]{trodden@physics.upenn.edu}
\affiliation{Center for Particle Cosmology, Department of Physics and Astronomy, University of Pennsylvania 209 S. 33rd St., Philadelphia, PA 19104, USA}

\begin{abstract}
We investigate the existence and behavior of oscillons in theories in which higher derivative terms are present in the Lagrangian, such as galileons. Such theories have emerged in a broad range of settings, from higher-dimensional models, to massive gravity, to models for late-time cosmological acceleration. By focusing on the simplest example---massive galileon effective field theories---we demonstrate that higher derivative terms can lead to the existence of completely new oscillons (quasi-breathers). 
We illustrate our techniques in the artificially simple case of 1 + 1 dimensions, and then present the complete analysis valid in 2 + 1 and 3 + 1 dimensions, exploring precisely how these new solutions are supported entirely by the non-linearities of the quartic galileon. These objects have the novel peculiarity that they are of the differentiability class $C^1$.
\end{abstract}

\maketitle

\section{Introduction}

Essentially all of modern physics is described by non-linear partial differential equations (PDEs). In many cases, it is sufficient to study highly symmetric configurations such that these reduce to ordinary differential equations or to study some linearized regimes in which the equations are easily solved. More generally, of course, multi-dimensional and non-linear processes are crucial to the understanding of a host of phenomena, from weather patterns, to black hole physics, to the early Universe. Among the rich and varied phenomena exhibited by non-linear PDEs, one class of particularly interesting objects not present in the linear regime is solitons, localized particle-like excitations that can be stable or extremely long-lived. Some examples of solitons include solitary water waves, domain walls, kinks, Skyrmions, and vortices. Typically, solitons can be classified into two classes: topological and non-topological solitons. The former are localized and stable due to the presence of non-trivial homotopy groups of the vacuum manifold of the field equations. These lead to defects that cannot relax to zero energy configurations due to some conserved topological quantity. The latter are localized and carry some conserved Noether charge $Q$ that requires an energetically unfavorable decay in order to remain conserved. {When the conserved charge results from a global $\mathrm{U}(1)$ symmetry (or an unbroken $\mathrm{U}(1)$ of some higher gauge group)} these objects are {typically} referred to as \emph{Q-balls} \cite{Coleman:1985ki}. 

Another class of {interesting non-linear solitary objects} are oscillons, sometimes called breathers. Oscillons are stable, extended, quasi-periodic (in time) particle-like excitations. These are objects with no conserved charges at all and for this reason are typically found in real scalar field theories. Their stability is due to {non-linearities that result in some of their comprising harmonics being localized, and unable to propagate to infinity}. In some cases, breathing objects may not be exact solutions, and so may be long-lived rather than absolutely stable. It is these objects, referred to as \emph{quasi-breathers}, that this paper is concerned with. Oscillons can occur in Bose-Einstein condensates and may be relevant in the early Universe. Indeed, the end state of many models of scalar field inflation is a Universe dominated by oscillons that form after the scalar condensate fractures due to some instability or through parametric reheating \cite{Amin:2010dc,Amin:2011hj}. Classical oscillons are supported by non-linear potentials that are \emph{shallower than quadratic} away from the minimum \cite{Amin:2010jq}, but oscillons have also been found in $P(\phi,X)$ theories (where $X=-(\partial\phi)^2/2$ is the canonical kinetic term) \cite{Amin:2013ika}, and, in particular, the Dirac-Born-Infeld (DBI) action can give rise to such objects. Oscillons can also exist in multi-field theories and may form after multi-field inflation driven by string moduli \cite{Antusch:2017flz}. They can also be formed from phase transitions and topological defects in the early universe \cite{Konoplich:1999qq,Khlopov:1999ys,Dymnikova:2000dy,Rubin:2000dq}.

The purpose of this paper is to determine the existence of oscillons in more general classes of higher derivative effective field theories \cite{Horndeski:1974wa,Nicolis:2008in,Deffayet:2009wt,Deffayet:2011gz,Zumalacarregui:2013pma,Gleyzes:2014dya,Gleyzes:2014qga,Langlois:2015cwa,Crisostomi:2016czh,Crisostomi:2016tcp,BenAchour:2016fzp}. We focus in particular on a simple representative of such theories, known as galileons \cite{Nicolis:2008in}, and perform a preliminary investigation of their properties. Galileon theories are a class of higher-derivative effective field theories whose equations of motion are precisely second-order and therefore are not plagued by Ostrogradski ghosts. Rather than being a fully covariant theory, they are defined on Minkowski space and as such they arise in the decoupling or low-energy limits of many different physical theories, including the Dvali-Gabadaze-Porrati (DGP) braneworld model \cite{Dvali:2000hr}, ghost-free Lorentz-invariant massive gravity \cite{deRham:2010kj,deRham:2014zqa}, and higher-derivative covariant theories constructed to avoid Ostrogradski ghosts \cite{Deffayet:2009wt,Deffayet:2011gz}. Our motivations for searching for breather solutions in galileon theories are manyfold. First, galileons may play a r\^{o}le in the early Universe.\footnote{Galileons were also a dark energy candidate \cite{Barreira:2014jha}, although their phenomenological viability is more tenuous after GW170817 \cite{Sakstein:2017xjx,Creminelli:2017sry,Baker:2017hug,Ezquiaga:2017ekz,Ezquiaga:2018btd} and cosmic microwave background (CMB) measurements \cite{Renk:2017rzu}.} Indeed, galileon inflation has been extensively studied (see \cite{Burrage:2010cu,Kobayashi:2010cm} for example) but the end point of this process is not clear. Finding galileon oscillons is one step towards determining if galileon breathers could be formed after inflation. Similarly, alternatives to inflation, in particular galilean genesis, utilizes galileon field theories \cite{Creminelli:2010ba,LevasseurPerreault:2011mw}. Another motivation is that massive galileons are a proxy for ghost-free massive gravity \cite{deRham:2017imi} and oscillons may act as a potential signature of these theories. Finally, the Vainshtein mechanism and non-renormalization theorems that galileon theories enjoy \cite{Nicolis:2008in,Goon:2016ihr} means that the higher-derivative operators are within the regime of validity of the effective field theory (EFT) in contrast to $P(\phi,X)$ theories. One example where this is problematic for solitons is Skyrmion theories where higher-derivative operators are required for stability so that such objects are outside the regime of validity of the EFT. For this reason, finding solitons in galileon theories is interesting in its own right. There is a no-go theorem for static solitons \cite{Endlich:2010zj} but the existence and stability of time-dependent solitons is still unexplored. 

\subsection{Summary of Results and Plan of the Paper} 

In this paper we perform a first investigation of oscillon solutions in massive galileon EFTs. Because the equations of motion resulting from higher-derivative theories are rather complicated, it is instructive to find the simplest system that encapsulates the relevant and new physics; as we shall see, galileons satisfy this requirement. The existence of galileon operators is dimension and geometry dependent. For this reason, we will analyze galileon breathers in $d$ + 1 dimensions where $d=1,\,2,\,3$ separately. We determine the existence of oscillons by looking for small-amplitude solutions controlled by some parameter $\ve<1$, which is used as an expansion parameter to construct an asymptotic series. Furthermore, we choose parameters such that the galileon operators are sufficiently large that their contribution to the equations of motion enters at the same order in $\ve$ as the canonical kinetic term and mass. (Another way of saying this is that they enter at a lower order in $\ve$ than one would n\"{a}ively expect.) This approach is akin to the construction of so called \emph{flat-top} oscillons \cite{Amin:2010jq}, where one guarantees that higher-order operators are important by taking some dimensionless parameter to be $\gg 1$. While this is necessary to find solutions analytically, one typically finds similar large-amplitude objects numerically for generic parameter choices \cite{Amin:2010jq}, and so this assumption can ultimately be relaxed. 

Our main results are as follows:
\begin{enumerate}
\item {\bf 1 + 1 dimensions}: Only the cubic galileon exists and it is unable to support quasi-breathers without the aid of other non-galileon operators. We give an example of this by including shift-symmetric, but not galileon symmetric, operators and find novel breather solutions. This example serves as a warm-up exercise designed to gain an analytic handle on the problem, and to glean insight into the construction of galileon oscillons before moving to higher dimensions, where the equations are less analytically tractable. We construct the profiles for these objects numerically and calculate their amplitude as a function of the model parameters.\clearpage
\item {\bf  3 + 1 dimensions}: We find novel oscillons/quasi-breathers supported by the non-linearity of the quartic galileon and the canonical kinetic term. These objects are solutions of a highly non-linear second-order ordinary differential equation. Surprisingly, and in contrast to other oscillons relevant for cosmology, these objects are $C^1$ functions and inhabit the space $W^{1,1}$.\footnote{We remind the reader that a function belongs to the differentiability class $C^k$ if its derivatives $f, f',\ldots, f^{(k)}$ exist and are continuous. A function belongs to the Sobolev space $W^{k,p}$ if its weak derivative up to order $k$ has a finite $L^p$ norm. The existence of $C^1$ oscillons is not unreasonable since the solution space of PDEs is indeed $W^{p,k}$ rather than $C^\infty$ (or even $C^2$). Indeed, the Hilbert space for the Schr\"{o}dinger equation is $H^k=W^{k,2}$.}   
\end{enumerate}

In the next section we give a brief introduction to oscillons/quasi-breathers in theories with non-linear potentials and $P(\pi,X)$ (with $X=-(\partial\pi)^2/2$) operators. In the subsequent section we introduce the massive galileon EFT and discuss some salient theoretical features. In section \ref{sec:HDO} we derive and present the main results of our work outlined above. We begin with the warm-up example of constructing galileon breathers in 1 + 1 dimensions before moving on to study the quartic galileon in 3 + 1 dimensions.  We conclude and discuss the implications of our findings as well as future directions in section \ref{sec:concs}. Our metric convention is the mostly positive one, and we work in units where $\hbar=c=1$. 

\section{Oscillons and Quasi-Breathers}
\label{sec:oscwarmup}

Oscillons are localized quasi-periodic excitations of scalar field theories whose existence is not due to a conserved Noether charge nor to non-trivial homotopy groups of the vacuum manifold \cite{Bogolyubsky:1976nx,Gleiser:1993pt}. In some cases, they are exact solutions and are therefore absolutely stable. In others, they may only be an asymptotic series labelled by some small parameter $\ve$ with a finite radius of convergence say $|\ve|\le\ve_0$. It is common that the radius of convergence is zero and in these cases the objects are referred to as \emph{quasi-breathers}, since they eventually decay through radiation but are often long-lived. We will use both terms interchangeably in what follows. As an example, consider the theory in $D = d + 1$ dimensions \cite{Fodor:2008es,Amin:2013ika}
\begin{equation}\label{eq:wu1}
S=\int\dd^{d+1}x\left[X-\xi{X^2}+\cdots -\frac{1}{2}m^2\pi^2 -\frac{\lambda_3}{3}\pi^3\-\frac{\lambda_4}{4}\pi^4+\cdots\right],
\end{equation}
where the ellipsis denotes higher derivative and higher order terms in the EFT. This can be brought into dimensionless form by rescaling $\pi\rightarrow m^{(d-1)/2}\pi$, $x^\mu\rightarrow x^\mu/m$,  $\xi\rightarrow \xi/m^{d+1}$, $\lambda_3\rightarrow  \lambda_3m^{(5-d)/2} $, and $\lambda_4\rightarrow \lambda_4 m^{d-3}$, to yield
\begin{equation}\label{eq:actwarmup}
S=\int\dd^{d+1}x\left[X-\xi{X^2}+\cdots -\frac{1}{2}\pi^2 -\frac{\lambda_3}{3}\pi^3-\frac{\lambda_4}{4}\pi^4+\cdots\right],
\end{equation}
where all fields and coupling constants are now dimensionless. We are looking for quasi-periodic localized solutions and so we will assume a solution in the form of the asymptotic series whose radius of convergence is governed by some small parameter $\ve\ll1$. We anticipate objects whose characteristic size decreases with increasing $\ve$ and so we define the new spatial coordinates
\begin{equation}\label{eq:rhodef1}
\zeta^i=\ve x^i,\quad i=1,\,\ldots,\, d.
\end{equation}
We will be interested in homogeneous solutions with an internal $\mathrm{SO}(d)$ symmetry and so we also define
\begin{equation}\label{eq:rhodef2}
\rho^2=\eta^{ij}\zeta_i\zeta_j=\ve^2\eta^{ij}x_ix_j.
\end{equation}
Because we expect that the non-linearities will shift the characteristic frequency away from 1 ($m$ if one puts the dimensions back in), we define a new time coordinate
\begin{equation}\label{eq:taudef}
\tau=\omega(\ve)t,\quad \omega(\ve)=1+\sum_{j=1}^\infty\ve^j\omega_j ,
\end{equation}
for constants $\omega_j$. It is convenient to choose initial conditions and use the freedom of shifting the time coordinate to set
\begin{equation}\label{eq:omdef}
\omega=\sqrt{1-\ve^2} ,
\end{equation}
which we will assume from the outset. Note that, by construction, the frequency {of the fundamental harmonic} always gets shifted to values smaller than the natural frequency {so that the oscillon is primarily supported by harmonics with frequencies smaller than the field's mass (recall that higher harmonics are suppressed by powers of $\ve$)}. This ensures that {the oscillon is stable on a long time-scale since these modes are unable to propagate}. {The higher harmonics (with frequencies greater than the field's mass) can propagate and ultimately lead to radiation at infinity \cite{Segur:1987mg,boyd1,Fodor:2008du,Fodor:2009kf}}. \footnote{Mathematically, the oscillon will radiate  because the asymptotic expansion does not converge for any value of $\ve$, and so an oscillon formed from some initial data will radiate as it relaxes to the true solution. (Exceptions to this are integrable theories where the oscillon is an exact solution such as the sine-Gordon model.) The radiation occurs with a highly-suppressed rate because the frequencies of the Fourier modes that comprise the oscillon are $\mathcal{O}(\ve)$ whilst those of the outgoing radiation modes are $\mathcal{O}(1)$ \cite{Hertzberg:2010yz}. Oscillons may also radiate due to quantum effects \cite{Hertzberg:2010yz} or the effects of cosmological backgrounds \cite{Farhi:2007wj}.} Had we instead chosen conditions such that the frequency is shifted to higher values, the object would be unstable and would rapidly decay to free radiation. 

Given the above considerations, we make the ansatz
\begin{equation}\label{eq:pians}
\pi(\tau,\rho)=\sum_{n=1}^\infty \ve^n\phi_n.
\end{equation}
The equation of motion resulting from \eqref{eq:actwarmup} up to $\mathcal{O}(\ve^3)$ in the new coordinates is
\begin{equation}
\ve\left(\ddot{\phi}_1+\phi_1\right)+\ve^2\left(\ddot{\phi}_2+\phi_2+\lambda_3\phi_1^2\right)+\ve^3\left(\ddot{\phi}_3+\phi_3+2\lambda_3\phi_1\phi_2+\lambda_4\phi_1^3+3\xi\dot{\phi}_1^2\ddot{\phi}_1\right)=0 ,
\end{equation}
where a dot denotes a derivative with respect to $\tau$ and $\nabla^2$ is the Laplacian operator defined with respect to $\zeta^i$. Equating the $\mathcal{O}(\ve)$ term to zero one finds that $\phi_1$ obeys the equation of a simple harmonic oscillator with unit frequency
\begin{equation}\label{eq:SHO}
\ddot{\phi}_1+\phi_1=0 ,
\end{equation}
so that with a suitable choice of initial conditions one has
\begin{equation}
\phi_1(\tau,\rho)=\so(\rho)\cos(\tau),
\end{equation}
for an (as yet) undetermined function $\Phi_1(\rho)$. Substituting this into the $\mathcal{O}(\ve^2)$ equation one finds
\begin{equation}
\ddot{\phi}_2+\phi_2=-\frac{1}{2}\Phi_1^2\left[1+\cos(2\tau)\right].
\end{equation}
The unique solution of this equation, imposing the initial conditions $\phi_2(0,\rho)=\dot{\phi}_2(0,\rho)=0$ (imposing these conditions is tantamount to shifting the $\tau$ origin), is
\begin{equation}
\phi_2=\frac{\lambda_3}{6}\Phi_1^2\left[-3+\cos(\tau)+\cos(2\tau)\right].
\end{equation}
At third order in $\ve$ we then find
\begin{equation}
\ddot{\phi}_3+\phi_3 = \left[\partial_\rho^2\Phi_1+\frac{(d-1)}{\rho}\partial_\rho\Phi_1-\Phi_1+\Delta\Phi_1^3\right]\cos(\tau)+[\cdots]\cos(2\tau)+[\cdots]\cos(3\tau),\label{eq:osc_prof_ex} 
\end{equation}
with
\begin{equation}
\Delta \equiv \xi-\lambda_4+\frac{10}{9}\lambda_3^2.
\end{equation}
Now, the term proportional to $\cos(\tau)$ is a resonance term i.e. it oscillates at the natural frequency. We therefore have a secular growth in $\phi_3$ unless the coefficient of this term vanishes. This gives us a non-linear equation for $\Phi_1(\rho)$ in terms of $\Delta$. Quasi-breathers can exist if this equation has a non-trivial solution. One can show \cite{Fodor:2008es,Amin:2013ika} that this is the case provided that $\Delta>0$. (We do not give the proof here since it is not relevant in what follows; we refer the interested reader to references \cite{Fodor:2008es,Amin:2013ika}). This procedure of equating each term at order $\ve^n$ to zero can be repeated \emph{ad infinitum} to build up the oscillon profile order by order by demanding that all secular resonance terms vanish \cite{Fodor:2008es}.

\section{Massive Galileon Effective Field Theory}
\label{sec:MGEFT}

Galileon-invariant scalar field theories are those whose actions are invariant under the galileon symmetry
\begin{equation}
\pi(x^\mu)\rightarrow \pi(x^\mu)+v_\mu x^\mu + c,
\end{equation}
with $c$ and $v_{\mu}$ constant. In four spacetime dimensions there are four operators that respect this symmetry \cite{Nicolis:2008in}:
\begin{align}\label{eq:ogaldef}
\mathcal{O}_n^{\rm gal}(\pi)&=\pi\Pi_{[\mu_1}^{\mu_1}\cdots\Pi^{\mu_{n-1}}_{\mu_{n-1}]};\quad n=2,\,3,\,4,\,5,
\end{align}
where $\Pi_{\mu\nu}=\partial_\mu\partial_\nu\pi$ and square brackets denote the trace of a tensor with respect to the Minkowski metric $\eta_{\mu\nu}$. Since these operators contain two, three, four, and five powers of the field they are referred to as the quadratic, cubic, quartic, and quintic galileons respectively. Individually, one has
\begin{align}
\mathcal{O}_2^{\rm gal}(\pi)&=\pi\Box\pi\\
\mathcal{O}_3^{\rm gal}(\pi)&=\pi\left([\Pi]^2-[\Pi^2]\right)\\
\mathcal{O}_4^{\rm gal}(\pi)&=\pi\left([\Pi]^3-3[\Pi][\Pi^2]+2[\Pi^3]\right).
\end{align}
We will not work with the quintic galileon in this work for reasons that we will discuss later. The Wilsonian effective action for a general galileon theory is then
\begin{equation}\label{eq:wilson}
S_{\rm W}=\int\dd^4 x\left[\sum_{n=2}^5c_i\frac{\mathcal{O}_n^{\rm gal}(\pi)}{\Lambda^{3(n-2)}}+\mathcal{L}_{\rm HD}(\partial^2\pi,(\partial^2\pi)^2,\ldots)\right],
\end{equation}
where $\mathcal{L}_{\rm HD}$ represents galileon-invariant higher-derivative operators acting on the field. Importantly, there is a powerful non-renormalization theorem that protects the galileon operators in the action (\ref{eq:wilson}) \cite{Nicolis:2008in,Hinterbichler:2010xn,Goon:2016ihr}. In particular, the coefficients $c_i$ are not corrected by loops but rather the coefficients of the operators appearing in $\mathcal{L}_{\rm HD}$ receive $\mathcal{O}(1)$ corrections. It is therefore consistent to choose to take the galileon operators alone as an effective field theory, since one can maintain a parametrically large separation of scales between these operators and those appearing in $\mathcal{L}_{\rm HD}$. 

Another nice feature of the galileon operators is the Vainshtein mechanism. Fluctuations  of the field about $\delta\pi$ some background $\pi^0(x^\mu)$ can be described by the effective action
\begin{equation}
S=\int\dd^4x Z^{\mu\nu}\partial_\mu\delta\pi\partial_\nu\delta\pi,
\end{equation}
where the kinetic matrix is
\begin{equation}
Z^{\mu\nu}=\eta^{\mu\nu} +b_3\frac{\Box\pi^0}{\Lambda^3}+b_4\frac{(\Box\pi^0)^2-(\nabla_\mu\nabla_\nu)^2}{\Lambda^6}+\cdots.
\end{equation}
This allows for three regimes: the linear regime where $\Box\pi^0/\Lambda^3\ll1$ i.e. the fluctuations behave as if they are free; the non-linear or Vainshtein regime where $\Box\pi^0/\Lambda^3\gsim1$ but $\partial^2/\Lambda^2\ll1$ so that non-linearities are important but quantum corrections are not; and the quantum regime where $\partial^2/\Lambda^2\gg1$. This hierarchy is reminiscent of the one appearing in GR where non-linearities are important below some scale (the Schwarzchild radius) but quantum corrections are only important at a much smaller scale, the Planck length. Indeed, one can define an analogous Vainshtein radius by $\Box\pi^0(r_{\rm V})\sim\Lambda^3$. The Vainshtein mechanism plays an important r\^{o}le in the phenomenology of galileon theories. In particular, new forces mediated by galileons are highly-suppressed inside the Vainshtein radius, allowing them to evade solar system tests of gravity \cite{Sakstein:2017pqi}. Galileon fluctuations move on the light cones of $Z^{\mu\nu}$ and one ubiquitously finds superluminal phase and group velocities \cite{Burrage:2011cr}. This, among other issues, has raised the question of whether galileons can be embedded into a Lorentz-invariant UV-completion \cite{Keltner:2015xda,Kaloper:2014vqa,Padilla:2017wth,Millington:2017sea}, a question which we will not discuss further here. As discussed by \cite{deRham:2017imi}, the technical obstructions to embedding the galileons into a Lorentz-invariant UV completion can be ameliorated if the galileon has a mass. Furthermore, the mass term only breaks the galileon symmetry softly (provided that the field only couples to other fields in a galileon-invariant manner) so that the non-renormalization theorem is not spoiled \cite{Goon:2016ihr}. The galileons as defined above arise in certain decoupling limits of other theories, most notably DGP braneworld gravity \cite{Dvali:2000hr} (and its generalizations \cite{deRham:2010eu}) and ghost-free massive gravity \cite{deRham:2010kj,deRham:2014zqa}. In the former case, the scalar plays the r\^{o}le of the brane-bending mode, while in the latter case the scalar is the helicity-0 mode of the massive graviton. Thus, massive galileons are naturally embedded in interacting massive spin-2 theories. 

Allowing a mass term for the galileon opens up the possibility of finding periodic quasi-breathers and so the action we will consider in this work is
\begin{equation}\label{eq:action1}
S=\int\dd^4 x\left(\sum_{n=2}^4c_i\frac{\mathcal{O}_n^{\rm gal}(\pi)}{\Lambda^{3(n-2)}}-\frac{1}{2}m^2\pi^2\right),
\end{equation}
with $c_2=-1/2$, and where $\Lambda$ and $m$ are constants with units of mass. We will consider the case $c_4>0$ so that the theory admits the Vainshtein mechanism \cite{Nicolis:2008in} (we also require $c_3>-\sqrt{3c_4/4}$). Such actions are theoretically well-motivated and have been considered in a number of works \cite{deRham:2017imi,Bellazzini:2017fep,deRham:2017xox}. 
In order to make connections with the existing oscillon literature it will sometimes be necessary to consider extending the action \eqref{eq:action1} to include a term
\begin{equation}\label{eq:action2}
\Delta S=\int\dd^4 x \frac{X^2}{\mathcal{M}^4},
\end{equation}
which must enter with a positive sign to ensure that the theory has a Lorentz-invariant UV-completion \cite{Adams:2006sv}. This action is not galileon-invariant (although it is shift-symmetric) and is under less control as an EFT compared with the action \eqref{eq:action1}, although such terms can appear alongside some galileon operators when integrating out the heavy radial mode of a $\mathrm{U}(1)$ complex scalar \cite{Burgess:2014lwa}. The presence of these terms are constrained by positivity bounds \cite{Bellazzini:2017fep} but we will not impose these here since we will only use the action \eqref{eq:action2} for illustrative purposes and will not draw physical conclusions from the results. We will only use \eqref{eq:action2} in 1 + 1 dimensions because in that case, as we will see presently, galileon oscillons cannot exist without it. We will view this case as an instructional exercise in order to learn how to construct galileon oscillons rather than as a theory to be taken seriously in its own right. In 2 + 1 and 3 + 1 dimensions the action \eqref{eq:action1} is sufficient to support oscillons and we will have no need for the action \eqref{eq:action2}.
  
\section{Higher Derivative-Supported Oscillons}
\label{sec:HDO}
One major difference between galileon operators and others that are known to support oscillons is that they are highly-sensitive to the number of spacetime dimensions. In particular, in $D$ dimensions there exist $D$ galileon operators (ignoring the tadpole) $\mathcal{O}_i^{\rm gal}(\pi)$ with $i=2,\ldots,D+1$ \cite{Nicolis:2008in}. Furthermore, the equation of motion for $\mathcal{O}_{n}^{\rm gal}(\pi)$ vanishes identically for configurations in which the galileon field depends on $n-1$ or fewer coordinates \cite{Bloomfield:2014zfa}. For example, the cubic galileon vanishes in 1 + 1 dimensions for static configurations and the quartic vanishes for cylindrical configuration in 2 + 1 dimensions. We are interested in $\mathrm{S}^{d-1}$-symmetric configurations in $D= d + 1$ dimensions and therefore we expect the cubic to contribute in 1 + 1 dimensions, the quartic and the cubic to contribute in 2 + 1 dimensions, and all galileon operators to contribute in 3 + 1 dimensions (although we will not study the quintic). It is useful to work with dimensionless quantities in what follows and so we rescale the coordinates and fields as
\begin{equation}
\tilde{x}^\mu=mx^\mu,\quad \pi=m^{\frac{d-1}{2}}\tilde{\pi}
\end{equation}
and define
\begin{equation}
 \xi \equiv \left(\frac{m}{\mathcal{M}}\right)^{d + 1},\quad g_3 \equiv c_3\left(\frac{m}{\Lambda}\right)^{\frac{d+3}{2}},\quad \textrm{and} \quad g_4 \equiv c_4\left(\frac{m}{\Lambda}\right)^{d+3},
\end{equation}
where we are now working in an arbitrary number of dimensions. The action we work with is then (dropping the tildes)
\begin{equation}\label{eq:galaction}
S=  \int \dd^{d + 1}x \left[\pi\Box\pi+\xi X^2 + \frac{g_3}{2}\pi\left([\Pi]^2-[\Pi^2]\right)+\frac{g_4}{24}\pi\left([\Pi]^3-3[\Pi][\Pi^2]+2[\Pi^3]\right) - \pi^2\right] ,
\end{equation}
with all quantities dimensionless. We will now look for oscillon solutions of \eqref{eq:galaction} following the procedure outlined in section \ref{sec:oscwarmup}. 

\subsection{Warm Up: 1 + 1 Dimensions}

As remarked above, only the cubic galileon contributes in 1 + 1 dimensions. We will make the ansatz \eqref{eq:pians} for $\pi$ and \eqref{eq:omdef} for $\omega$ and work in the $\rho(=\ve x)$ and $\tau$ coordinates defined in \eqref{eq:rhodef2} and \eqref{eq:taudef} respectively. By construction, $\phi_1$ satisfies equation \eqref{eq:SHO} so that 
\begin{equation}
\phi_1=\Phi_1(\rho)\cos(\tau).
\end{equation}
Now let us examine the equation of motion:
\begin{align}
\omega^2\ddot{\pi}-\ve^2\pi'' +
\pi&\nonumber+\xi\left[3\omega^2\dot{\pi}^2\ddot{\pi}-\ve^2\omega^2\pi'^2\ddot{\pi}-4\omega^2\ve^2\dot{\pi}\pi'\dot{\pi}'-\omega^2\ve^2\dot{\pi}^2\pi''+3\ve^4\pi'^2\pi''\right]\\&+g_3\ve^2\omega^2(\ddot{\pi}\pi''-\dot{\pi}'^2)=0.
\label{eq:cubic1p1eqn}
\end{align}
In order to obtain a non-trivial profile for $\Phi_1$ we need the cubic galileon terms (proportional to $g_3$) to contribute to the resonance term (proportional to $\cos(\tau)$) at order $\ve^3$. Now, the cubic galileon contribution to the equation of motion is quadratic in $\pi$ and so the solution for $\phi_1$ will not contribute to the resonance term since $\phi_1^2\sim\cos^2(\tau)$ can be expanded in terms of even harmonics only i.e. the expansion contains $\cos(2\tau)$ but not $\cos(\tau)$. On the other hand, the product $\phi_1\phi_2\sim\cos(\tau)\cos(2\tau)\sim\cos(\tau)+\cdots$ and will therefore contribute to the resonance term. Following the discussion in section \ref{sec:oscwarmup}, we therefore need this term to contribute to the $\mathcal{O}(\ve^3)$ equation of motion. This can be achieved if we take $g_3\sim \ve^{-2}$, i.e. we define our expansion parameter $\ve\sim1/\sqrt{g_3}$. We therefore define $g_3= \gbt/\ve^2$, which is tantamount to considering the region of parameter space where $m^3/\Lambda^3\gg1$. This is analogous to the procedure used to find analytic solutions to potential-supported theories, in which oscillon solutions are stabilized by $g\pi^6$ terms in the scalar potential, resulting in flat-top oscillons \cite{Amin:2010jq}. In that scenario, the contribution that would typically enter at $\mathcal{O}(\ve^5)$ enters at $\mathcal{O}(\ve^3)$ after one makes the choice $g\sim\ve^{-2}\gg1$. While this choice is necessary to find solutions analytically and to determine the existence of such objects with small amplitudes, numerical simulations reveal that large-amplitude objects with $g\sim\mathcal{O}(1)$ persist but cannot be found analytically. We expect similar features in galileon theories and so we do not treat $m^3/\Lambda^3\gg1$ as being a necessary condition for the existence of oscillons, but rather as a useful limit in which to find small-amplitude objects analytically. 

Returning to the calculation at hand, the second-order equation of motion is then
\begin{equation}
\ddot{\phi}_2+\phi_2+\gbt\left(\ddot{\phi}_1\phi_1''-\dot{\phi}_1'^2\right)=0.
\end{equation}
with solution 
\begin{equation}
\phi_2=\frac{\gbt}{2}\left(\so''\so+\so'^2\right)-\frac{\gbt}{3}\left(\so''\so+2\so'^2\right)\cos(\tau)+\frac{\gbt}{6}\left(\so'^2-\so''\so\right)\cos(2\tau),
\end{equation}
where we have imposed the initial condition $\phi_2(0,\rho)=\dot{\phi}_2(0,\rho)=0$. The order $\ve^3$ equation is then 
\begin{align}
\ddot{\phi}_3+\phi_3&=\cdots\nonumber\\&+ \left[\so''-\so+\frac34\so^3+\gbt^2\left(\frac23\so^2\so''+\frac54\so''^2\so+\frac53\so^{(3)}\so'\so+\frac{5}{12}\so^{(4)}\so^2\right)\right]\cos(\tau)\nonumber\\& +\left[\cdots\right]\cos(3\tau),\label{eq:cubicthridordereqn}
\end{align}
where the ellipsis corresponds to expressions that will not require the detailed form of in what follows. In order to have periodic solutions we must demand that the coefficient of the resonance term ($\cos(\tau)$) vanishes, which gives us an equation for the oscillon profile $\so(\rho)$
\begin{equation}
\so''-\so+\frac{3\xi}{4}\so^3+\gbt^2\left(\frac23\so^2\so''+\frac54\so''^2\so+\frac53\so^{(3)}\so'\so+\frac{5}{12}\so^{(4)}\so^2\right)=0.
\end{equation}
This equation has two parameters, $\xi$ and $\gbt$, but we can remove $\xi$ by scaling $\so\rightarrow\so/\sqrt{\xi}$ and $\gbt\rightarrow\gbt\sqrt{\xi}$ and so we can use this scaling to fix $\xi=1$. We will do this presently but for now it is instructive to keep $\xi$ free. Multiplying by $\so'$ one finds that this equation has a first integral or \emph{conserved energy density} given by
\begin{equation}\label{eq:firstintcubic}
\mathcal{E}=\frac12\so'^2-\frac12\so^2+\frac{3\xi}{16}\so^4+\gbt^2\left(\frac56\so''\so'^2\so-\frac{1}{24}\so'^4-\frac{5}{24}\so''^2\so^2+\frac{5}{12}\so^{(3)}\so'\so^2\right).
\end{equation}
Now, we are looking for objects that satisfy the free (linear) equation of motion ($\so''-\so=0$) at large distances i.e.
\begin{equation}\label{eq:rholarge1d}
\lim_{\rho\rightarrow\pm\infty}\so(\rho)=\mathcal{B}_1e^{\pm\rho} ,
\end{equation}
(the constant $\mathcal{B}_1$ must be solved by matching onto the boundary conditions at the origin) but are localized near the origin due to non-linear self-interactions. This configuration has zero conserved energy and we are therefore looking for solutions with $\mathcal{E}=0$. One can find a further condition by demanding that the profile is symmetric about the origin ($\so'(0)=\so^{(3)}(0)=0$):
\begin{equation}\label{eq:origincubic}
\so''(0)=-\sqrt{\frac{3\xi}{10g^2}}\sqrt{3\xi\so(0)^2-8}.
\end{equation}
This gives us a relation between the amplitude and the second derivative at the origin. Note that a second solution with $\so''(0)>0$ exists, but we have discarded it since it would give rise to an increasing function away from the origin. Such solutions are typically higher energy and are unstable. One can see from equation \eqref{eq:origincubic} that the $X^2$ term in the action is necessary in order to have a galileon supported oscillon. Indeed, had it been absent then the second derivative would be imaginary so that no oscillon solutions could exist. Including it allows for oscillon solutions provided that
\begin{equation}
\so(0)^2>\frac{8}{3\xi}.
\end{equation}
If one takes $\so(0)^2=\frac{8}{3\xi}$ then the profile is flat. Interestingly, the amplitude for $P(\pi,X)$-supported oscillons is precisely $\sqrt{8/3\xi}$ \cite{Amin:2013ika} so that galileon oscillons necessarily have larger amplitudes than $P(\pi,X)$-oscillons for fixed parameters.

\begin{figure}
\centering
\includegraphics[width=0.45\textwidth]{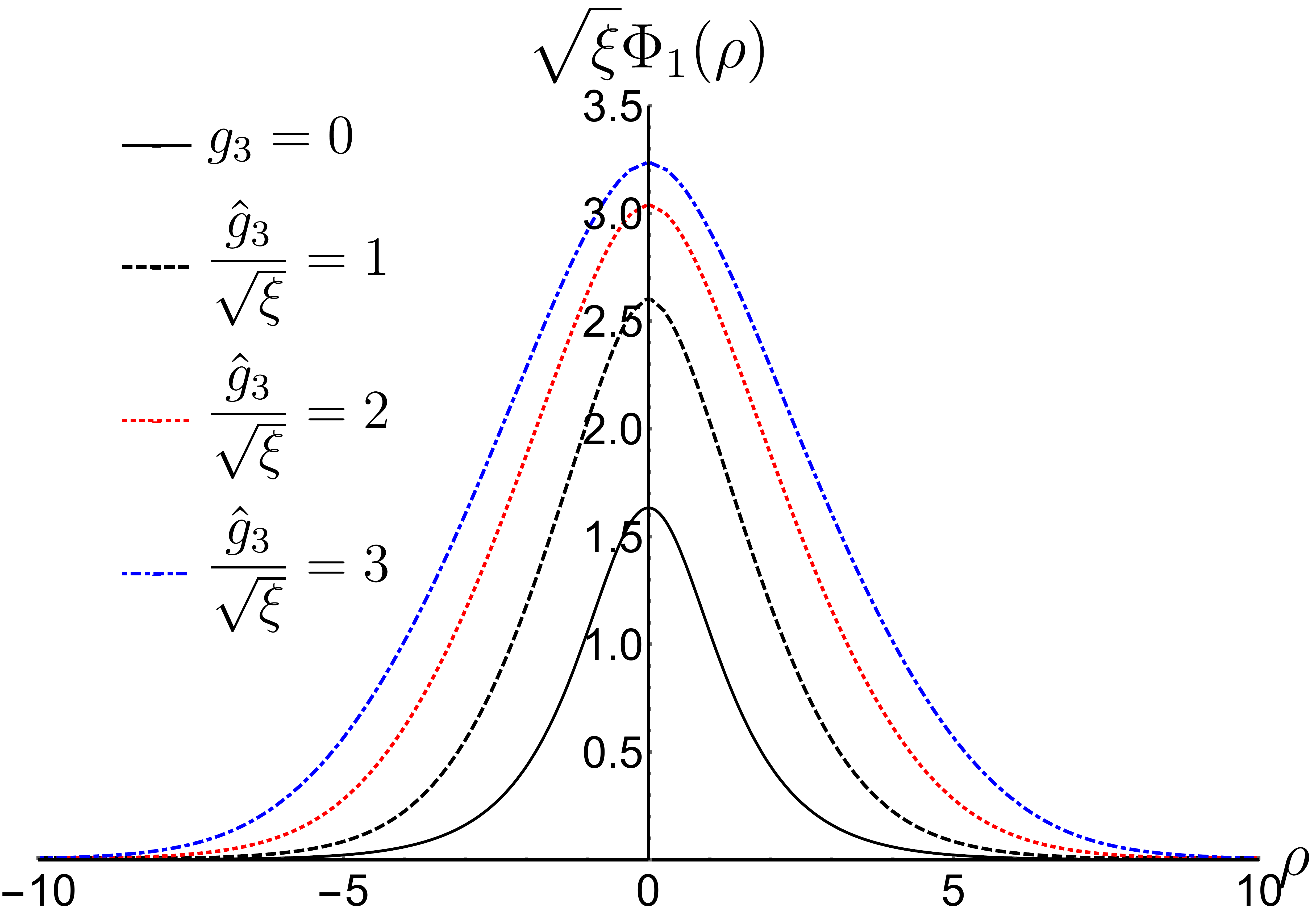}
\includegraphics[width=0.45\textwidth]{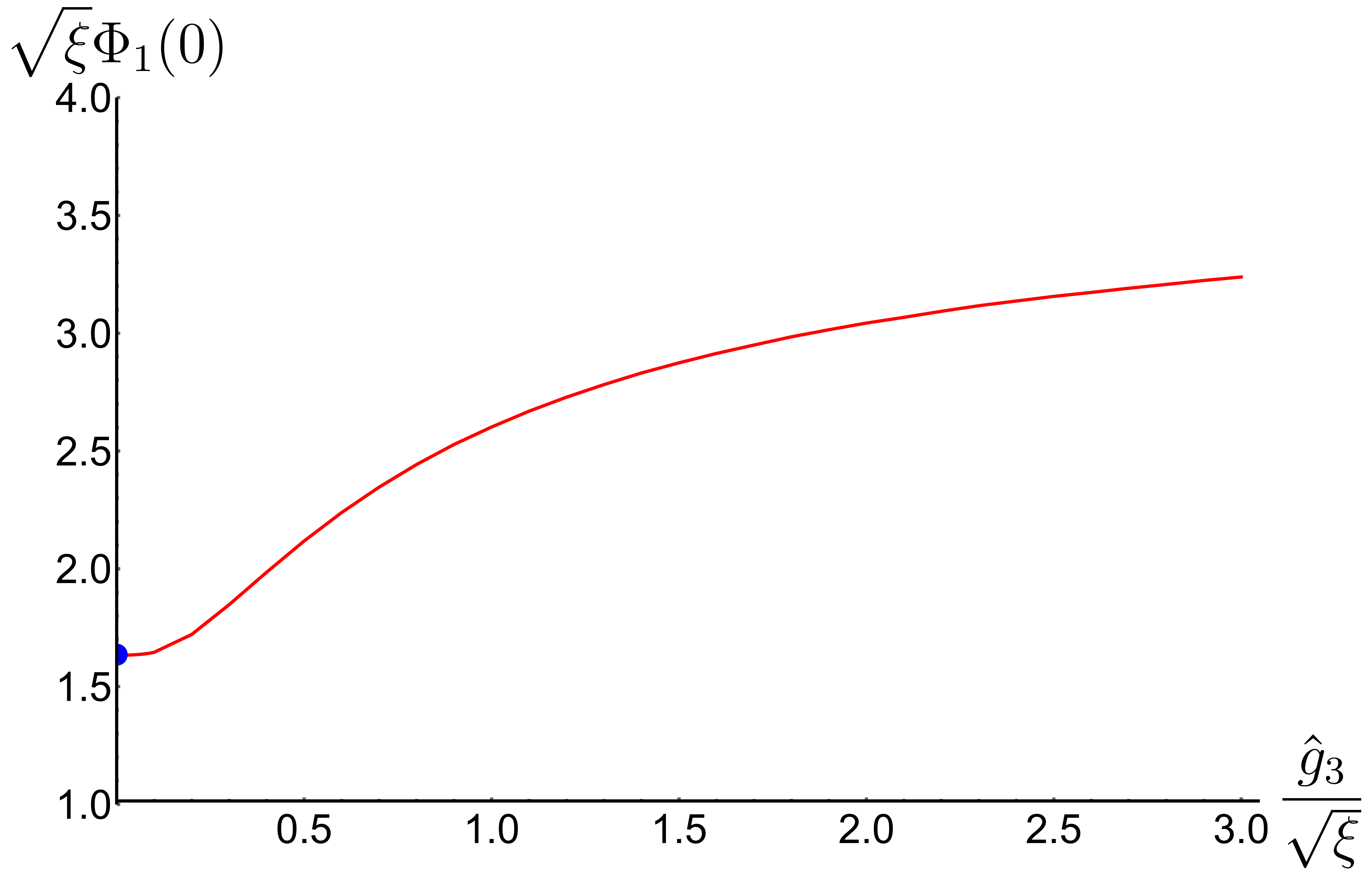}
\caption{\emph{Left}: The oscillon profile for $\gbt/\sqrt{\xi}=1$ (black, dashed), $\gbt/\sqrt{\xi}=2$ (red, dotted), and $\gbt/\sqrt{\xi}=3$ (blue, dot-dashed). The black line corresponds to the $P(\pi,X)$-oscillon profile. \emph{Right}: The amplitude $\so(0)$ as a function of $\gbt/\sqrt{\xi}$. The $P(\pi,X)$ prediction $\sqrt{\xi}\so(0)=\sqrt{8/3}$ corresponds to the blue point.}\label{fig:wcubic}
\end{figure}

Unlike the case of potential or $P(\pi,X)$-supported oscillons \cite{Fodor:2008es,Amin:2010jq,Amin:2013ika}, the equation governing the profile ($\mathcal{E}=0$ in equation \eqref{eq:firstintcubic}) does not have an analytic solution and so we must proceed numerically. We are looking for solutions that are localized near the origin, so that non-linear terms are important, but that tend to the linear solution $\exp(-\rho)$ at large distances (see equation \eqref{eq:rholarge1d}). The task at hand is then to solve equation \eqref{eq:firstintcubic} (with $\mathcal{E}=0$) given the boundary conditions $\so(0)=\so^{(3)}(0)=0$ and \eqref{eq:origincubic}. This leaves the value of $\so(0)$ undetermined and so, in the current formulation, the correct solution must be found by solving the equation for different values of $\so(0)$ such that $\lim_{\rho\rightarrow\infty}\so(\rho)\sim\exp{(-\rho)}$. This is a time-intensive process but it can be simplified dramatically by reformulating the problem in terms of the phase space variables $\{\so(\rho),\so'(\rho)\}$. We give the technical details of this process for the interested reader in Appendix \ref{app:cubic}. Some examples of the oscillon profiles for varying $\gbt/\sqrt{\xi}$ (the only free combination of parameters)) are given in the left panel of fig. \ref{fig:wcubic}; one can see that galileons produce oscillons with similar shapes, and with larger amplitudes that are increasing functions of $\gbt/\sqrt{\xi}$. The right panel of fig. \ref{fig:wcubic} shows the amplitude $\sqrt{\xi}\so(0)$ as a function of $\gbt/\sqrt{\xi}$. One can see that it is indeed an increasing function. In the case of $P(\pi,X)$ oscillons, the profile was calculated analytically in \cite{Amin:2013ika} as
\begin{equation}\label{eq:kes}
\so(\rho)=\sqrt{\frac{8}{3\xi}}\sech(\rho),
\end{equation}
which is also shown in both figures. In the case of the right panel, the amplitude is shown using the blue point. Evidently, the amplitude tends to this value for small $\gbt/\sqrt{\xi}$. One interesting difference between galileon oscillons and oscillons in $P(\pi,X)$ theories is that the boundary conditions do not impose any limit on the amplitude. One cannot have arbitrarily large amplitudes whilst simultaneously satisfying the boundary conditions in $P(\pi,X)$ theories \cite{Amin:2010jq,Amin:2013ika}.

\subsection{3 + 1 Dimensions}
\label{sec:3+1}

In three spatial dimensions there are contributions from the cubic, quartic and quintic galileon operators. As remarked above, we will show presently that the combination of the kinetic term and the quartic galileon admits quasi-breather solutions, and so we will set $\xi=0$ from here on, having no other justification for incorporating it into the massive galileon EFT. Furthermore, the lessons we have learned from our warm up exercise in 1 + 1 dimensions give us good cause to neglect the cubic and quintic terms too. Recall that the cubic galileon contributed terms of order $\ve^2g_3\phi^2$ (ignoring time and space derivatives) to the equation of motion, which forced us to make a suitable choice of scaling for $g_3$ ($g_3\sim \ve^{-2}$) to ensure that this term contributed to the $\mathcal{O}(\ve^2)$ equation of motion and therefore that terms such as $\phi_1\phi_2$ (again, suppressing derivatives) appeared in the $\mathcal{O}(\ve^3)$ equation. This was necessary to ensure that the cubic galileon contributed to the resonance term (proportional to $\cos(\tau)$). Had we not made this choice and had instead chosen $g_3$ such that the cubic operator contributed at third (and higher) order, the quadratic nature ($g_3\phi_1^2$) of the equation of motion would mean that no odd harmonics were present. In fact, this is completely analogous to the cubic potential discussed in section \ref{sec:oscwarmup}, the contribution to the EOM is quadratic in the field and therefore the above process is necessary. The only difference is that we had to take $g_3\sim \mathcal{O}(\ve^{-2})$ owing to the higher-derivative nature of the galileons, whereas one can take $\lambda_3\sim\mathcal{O}(1)$ for potential-supported oscillons. 

Now, the contribution of the quartic galileon to the equation of motion is cubic in the fields and hence it is sufficient to choose $g_4$ to scale with $\ve$ in such a way that it first contributes at order $\ve^3$ because terms such as $\phi_1^3$ (suppressing derivatives) contribute odd harmonics, including the resonance term $\cos(\tau)$. This is analogous to the quartic potential discussed in section \ref{sec:oscwarmup}, the difference being that $g_4$ must be chosen to cancel the effects of higher-derivatives whereas $\lambda_4\sim\mathcal{O}(1)$. 

Let us now briefly discuss the quintic galileon. This contributes quartically to the equation of motion and so the situation is akin to the cubic rather than the quartic: one must choose $g_5$ to scale with $\ve$ such that the quintic contributes to both the $\mathcal{O}(\ve^2)$ and $\mathcal{O}(\ve^3)$ equations in order for the resonance term to be affected by its presence. Given the above considerations, we will not include the cubic galileon in what follows since it greatly complicates the equations in 3 + 1 (and 2 + 1) and all of the new and salient features are captured by including the quartic solely.  Similarly, we will not discuss the quintic galileon at all in this paper. One can forbid these terms either by imposing a $\mathbb{Z}_2$ symmetry or the symmetry of the special galileon \cite{Hinterbichler:2015pqa,Novotny:2016jkh}.

The equation of motion in the $\tau$--$\rho$ coordinate system is 
\begin{align}
\omega^2\ddot{\pi}&-\ve^2\left(\pi''+\frac{2}{\rho}\pi'\right) +
\pi
+\ve^4g_4\frac{\pi'}{\rho}\left[\frac{\omega^2\ddot{\pi}\pi'}{\rho}-\ve^2\frac{\pi''\pi'}{\rho}+2\omega^2\ddot{\pi}\pi''-\omega^2\dot{\pi}'^2\right]=0.
\label{eq:3deqn}
\end{align}
As per our discussion above, we must choose $g_4$ such that the quartic contributes at order $\ve^3$, and so we choose $g_4= \hat{g}_4/\ve^4$. (One can equivalently view this as the definition of our small number $\ve\sim g_4^{-1/4}$.) Once again we need to employ a procedure similar to the flat-top oscillon construction, whereby we push the contributions of higher-order operators to lower-order in $\ve$ by taking a dimensionless parameter to be $\gg1$. In this case, this implies we are in the regime where $m^6/\Lambda^6\gg1$. As discussed above, we do not take this as a necessary condition, but rather as a tool that allows us to construct solutions analytically. We expect similar objects to exist for other parameter choices, the difference being that they must be found numerically. 

Following the procedure in \ref{sec:oscwarmup}, the order $\ve$ and $\ve^2$ equations of motion are
\begin{align}
\ddot{\phi}_1+\phi_1&=0\Rightarrow\phi_1=\so(\rho)\cos(\tau)\label{eq:phi_13+1}\\
\ddot{\phi}_2+\phi_2&=0\Rightarrow\phi_2=0,\label{eq:phi_13+12}
\end{align}
where we have set $\phi_2=0$ using appropriate boundary conditions. Using equation \eqref{eq:phi_13+1} in equation \eqref{eq:3deqn} one finds the third-order equation of motion
\begin{align}
\ddot{\phi}_3+\phi_3&=\cdots\nonumber\\&+ \left[\so''+\frac{2}{\rho}\so'-\so+\frac{\gbf}{2\rho}\left(\so'^3+3\so\so'\so''+\frac{3\so\so'^2}{2\rho}\right)\right]\cos(\tau)\nonumber\\& +\left[\cdots\right]\cos(3\tau),\label{eq:3+1thridordereqn}
\end{align}
where we have once again given only the coefficient of the resonance term. This must be identically zero in order to avoid secular growth. Scaling $\so\rightarrow\so/\sqrt{\hat{g}_4}$ one then finds the equation governing the oscillon profile:
\begin{equation}\label{eq:3+1profeqn}
\left(1+\frac{3\so'\so}{2\rho}\right)\so''+\frac{\so'^3}{2\rho}+\frac{3\so\so'^2}{4\rho^2}+\frac{2}{\rho}\so'-\so=0.
\end{equation}
The task of finding oscillon solutions is then to solve this equation given the boundary condition\footnote{Imposing this, equation \eqref{eq:3+1profeqn} gives \[\so''(0)=-\frac{2}{3\so(0)}\left(1\pm\sqrt{1+\so(0)^2}\right)\] so there is no restriction on $\so(0)$ or $\so''(0)$ as there was in 1 + 1 dimensions.} $\so'(0)=0$ with $\so(0)$ chosen such that 
\begin{equation}\label{eq:3dlim}
\lim_{\rho\rightarrow\infty}\so(\rho)\sim \mathcal{B}_3\frac{e^{-\rho}}{\rho}
\end{equation}
for some constant\footnote{This constant is not arbitrary because the full equation \eqref{eq:3+1profeqn} is non-linear. The value of $\mathcal{B}_3$ is a global property of the solution i.e. it depends on $\so(0)$.} $\mathcal{B}_3$ i.e. $\so$ is the spherically-symmetric solution of the linear equation $\nabla^2\so-\so=0$ (in three spatial dimensions) at large distances. Let us recall how this is accomplished for $P(\pi,X)$ and potential-supported oscillons in $d$ + 1 dimensions with $d>1$. Unlike in 1 + 1 dimensions, there is no conserved first integral\footnote{Of course, we still have energy and momentum conservation resulting from the Poincar\`{e} symmetries of the action but it is not necessarily the case that these should hold order by order in $\ve$.} and so one writes the equivalent of equation \eqref{eq:3+1profeqn} in the form
\begin{equation}\label{eq:Edrop}
\frac{\dd\mathcal{E}}{\dd\rho}=-\frac{2}{\rho}(\partial_\rho\so)^2,
\end{equation}
and uses the phase space approach. In particular, since the \emph{energy} $\mathcal{E}$ would be conserved if not for the right hand side, one can deduce that there are a series of discrete solutions with $\mathcal{E}(\rho=0)>0$ such that the phase space trajectories move from $(\so,\,\so')=(\so(0),0)$ to $(\so,\,\so')=(0,\,0)$ corresponding to $\lim_{\rho\rightarrow\infty}\mathcal{E}=0$, which is necessary to ensure that the solution tends to the linear one (i.e. the profile tends to the one given in equation \eqref{eq:3dlim}) at large distances. (We refer the reader to \cite{Anderson:1971pt} for the technical analysis of the phase space.) These solutions are characterized by the number of nodes in the profile, the lowest energy solution having zero nodes and higher energy solutions having an increasing number. Equation \eqref{eq:3+1profeqn} is not amenable to such an analysis for several reasons. First, although one can make judicious integrations by parts to reformulate it in the form of equation \eqref{eq:Edrop}, this is not useful because the energy depends on $\rho$, greatly complicating the phase space analysis. Furthermore, the right hand side contains a term of indefinite sign so that one cannot deduce anything meaningful about the energy along oscillon trajectories in phase space. Finally, the coefficient of $\so''$ can vanish identically at a point $\rho_s$ that depends on the boundary conditions, either at the origin or infinity depending on the direction from which $\rs$ is approached. At this point, the second derivative is not determined from the equation. This implies that solutions may not be smooth since one could construct functions that are discontinuous at $\rs$. For these reasons, it is instructive to proceed by analyzing the equation directly.

We begin by showing that solutions with nodes cannot exist. Solving equation \eqref{eq:3+1profeqn} for $\so''$ one finds
\begin{equation}\label{eq:pdd3+1g0}
\so''=\frac{4\rho^2\so-8\rho\so'-4\so\so'^2-2\rho\so'^3}{2\rho\left(2\rho+3\so\so'\right)}.
\end{equation}
Since $\so'(\rho)=0$ at any stationary point where $\rho\ne\{0,\,\infty\}$, one has $\so''(\rho)=\so(\rho)$ at any potential node. This equation implies that the stationary point is necessarily a minimum if $\so(\rho)>0$ and a maximum if $\so(\rho)<0$. Clearly this precludes the possibility that such a point is a node.  
In practice, we have not been able to find any smooth zero node solutions for reasons that we now discuss. 

Consider the point $\rho_s$ mentioned above where the coefficient of $\so''(\rho)$ in equation \eqref{eq:3+1profeqn} vanishes. This point is defined implicitly by $3\so'(\rs)\so(\rs)=2\rs$, and the second derivative $\so''(\rho)$ is undetermined there. To see this, focus on a solution that satisfies the boundary conditions at the origin, i.e. $\so'(0)=0$ for some $\so(0)$, and assume that 
\begin{equation}
\lim_{\rho\rightarrow{\rs}^-}\left(1+\frac{3\so'(\rho)\so(\rho)}{2\rho}\right)=\mathcal{A} .
\end{equation}
Then, equation \eqref{eq:3+1profeqn} with $\so'(\rs)=2\rs/(3\so'(\rs))$ gives
\begin{equation}
\so(\rs)=\pm\frac{1}{\sqrt{2}}\sqrt{-1\pm\sqrt{9-\frac{16\rho_s^2}{3}}} ,
\end{equation}
when $\mathcal{A}=0$, so that there are no solutions. If instead one takes $\mathcal{A}\ne0$ then real solutions do exist\footnote{It is tempting to conclude from the present discussion that it therefore follows that
\begin{equation*}
\lim_{\rho\rightarrow{\rs}^-}\so''(\rho) = \infty
\end{equation*}
but such a conclusion would be erroneous. Rather, the solution to equation \eqref{eq:3+1profeqn} should be viewed as a weak solution and is therefore locally integrable. One can only make meaningful statements about this function when integrated against tests functions.}, although we do not give them here since they are solutions of a more complicated quartic equation and their expressions are long and cumbersome. The same conclusions are reached if one begins with a solution satisfying equation \eqref{eq:3dlim} at $\rho\rightarrow\infty$ and lets $\rho\rightarrow{\rs}^+$. 

\begin{figure}
\centering
{\includegraphics[width=0.45\textwidth]{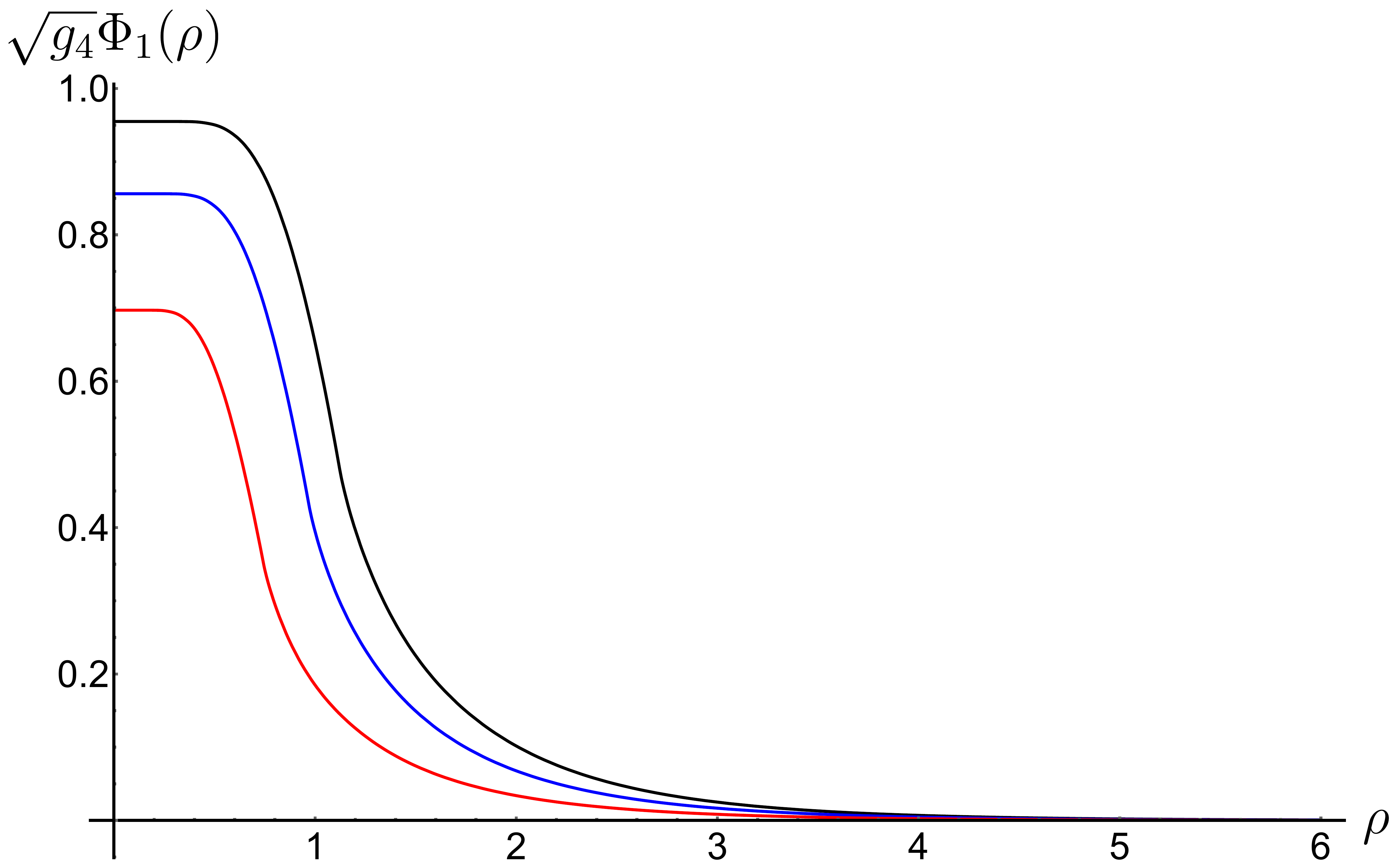}}
{\includegraphics[width=0.45\textwidth]{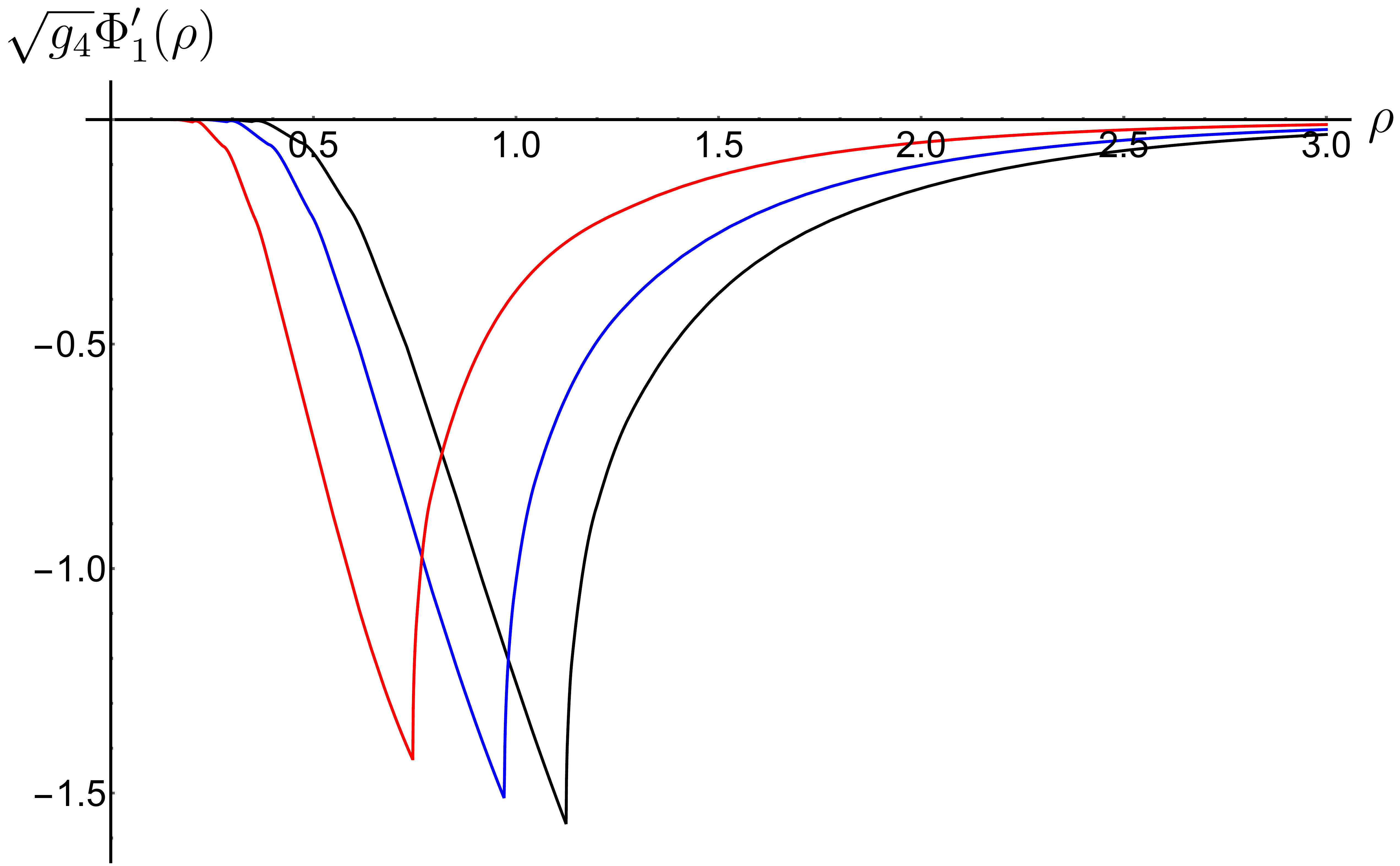}}
\caption{\emph{Left Panel}: Three profiles for different amplitude cubic galileon oscillons in 3 + 1 dimensions. \emph{Right Panel}: The first derivatives $\so'(\rho)$ of the profiles in the left panel. The colors in each figure correspond to the same solution.  }\label{fig:3Dprofs}
\end{figure}

We seek solutions over the entire positive real interval. Such solutions must therefore be of the smoothness class $C^1$. This may be surprising at first but we remind the reader that the solutions of partial differential equations naturally lie in Sobolev spaces rather than the space of $C^\infty$ (or even $C^2$) functions. In fact, it is not uncommon for such solutions to arise in non-linear wave equations, see e.g. \cite{1078-0947_2018_3_1567}. 
$C^1$ solutions of equation \eqref{eq:3+1profeqn} that span the positive real interval, $\so(\rho)\in W^{1,1}(\mathbb{R}^+)$, can be constructed numerically. One finds a continuous spectrum of solutions distinguished by their amplitudes $\so(0)$. Some examples are shown in figure \ref{fig:3Dprofs}. The left panel shows oscillons with different amplitudes. Evidently, these are very flat objects and are completely smooth. The flatness was anticipated, since we chose $g_4\gg1$, putting us in the flat-top regime as discussed above. The right panel shows $\so'(\rho)$. One can see that there is indeed a point $\rs$ where the derivative is continuous but not smooth, indicating the $C^1$ nature of the solution. We have verified numerically that $3\so'(\rs)\so(\rs)=2\rs$ at this point, as per our analytic prediction above.

\section{Conclusions and Outlook}
\label{sec:concs}

The study of higher-derivative effective field theories is important for a multitude of reasons. The leading-order derivative corrections to general relativity are higher-derivative in nature. Similarly, many infra-red modifications of gravity make use of higher-derivative structures, an important example being Horndeski gravity \cite{Horndeski:1974wa,Deffayet:2011gz} (and beyond \cite{Gleyzes:2014qga,BenAchour:2016fzp}) which can be viewed as a  framework for constructing ghost-free dark energy or inflation scenarios. Similarly, the decoupling limit of many IR modifications of gravity such as ghost-free massive gravity or braneworld models are described by higher-derivative EFTs, in particular massless galileons \cite{Nicolis:2008in,deRham:2010kj,Hinterbichler:2011tt,deRham:2014zqa}. 

Galileons have two remarkable features. First, they enjoy a non-renormalization theorem whereby the coefficients of the galileon operators are not corrected by self-loops or matter loops (provided that the coupling is galileon-invariant) \cite{Nicolis:2008in,Hinterbichler:2010xn,Goon:2016ihr}. Second, The Vainshtein mechanism suppresses fifth-forces compared with the gravitational force and ensures that the range of validity of the EFT extends to scales far smaller than one would n\"{a}ively expect. In particular, there is a \emph{Vainshtein radius}, inside which galileon non-linearities can be important while other higher-derivative operators are still suppressed. The range of validity is then lowered to smaller distances where these are comparable to the galileon terms. Interestingly, these features are not disrupted by the inclusion of a mass term for the galileon, even though it breaks the galileon symmetry \cite{Goon:2016ihr,deRham:2017imi,deRham:2017xox}. Massive galileon theories are more naturally embedded in ghost-free massive gravity than their massless counterparts and may shed some light on the nature of any UV-completion \cite{deRham:2017xox}.

Phenomenologically, the Vainshtein mechanism is incredibly efficient and ensures that galileon theories are difficult to test experimentally \cite{deRham:2012fg,Sakstein:2017pqi,Dar:2018dra} and one must resort to exotic and non-traditional tests of gravity \cite{Sakstein:2017bws,Sakstein:2015zoa,Sakstein:2015aac,Burrage:2017qrf,Adhikari:2018izo}. A powerful no-go theorem \cite{Endlich:2010zj} prohibits the existence of static topological soliton configurations within the regime of validity of the galileon EFT (this is an extension of Derrick's famous no-go theorem for potential-supported objects) but the fact that a small mass for the galileon is both theoretically well-motivated and interesting opens up the possibility of finding quasi-periodic localized solutions: oscillons (or quasi-breathers).

In this paper we have shown that massive galileon effective field theories can indeed result in quasi-breathers/oscillons.  In 1 + 1 dimensions the cubic galileon, which is the only galileon operator that can exist, can give rise to a new type of quasi-breather, provided that the EFT contains shift-symmetric (but not galileon-invariant) operators. In particular, the solution we found required the presence of the operator $X^2$ where $X\sim(\partial\pi)^2$. We analyzed the properties of these new objects and derived their amplitude as a function of the cubic galileon coupling. In 3 + 1 dimensions we found new quasi-breathers that are supported entirely by the non-linearities of the quartic galileon. These objects have the novel feature that they are $C^1$ functions.   

Having found these objects, several questions about their nature naturally arise. First, are they stable? Oscillons supported by non-linear scalar potentials are stable in 1 + 1 dimensions, but broad, small-amplitude oscillons are unstable to long wavelength perturbations in 3 + 1 dimensions \cite{Amin:2010jq} (large-amplitude objects are robust in 3 + 1 dimensions). Similarly, small-amplitude oscillons supported by $P(\pi,X)$ terms exhibit a long wavelength instability in $d\ge3$. In both of these cases, one can apply the Vakhitov-Kolokolov stability criterion \cite{1973R&QE...16..783V} to study the dynamics of long wavelength perturbations. The criterion assumes that the spatial kinetic term for perturbations is the d-dimensional Laplacian, which is not the case for galileon theories. One could attempt to generalize the criterion but that is beyond the scope of the present paper. Given the complexity of the field equations, it may be simpler to study the stability numerically, although this too is difficult given the higher degree of non-linearity in the equations \cite{Li:2013tda,Dar:2018dra}.

Another related question is the radius of convergence for the asymptotic expansion. Quasi-breathers radiate at wave modes $k=nm$ with $n=2\,,3,\,4,\,\ldots$ as a result of the radius of convergence (in terms of $\ve$) being zero on the real line \cite{Segur:1987mg,Fodor:2008du,Fodor:2009kf}. Said another way, oscillons radiate since they are not exact solutions of the field equations. Their long-lived nature is due to the fact that they are comprised of wave modes that are far smaller than those into which they radiate, and so the radiation process takes a long time \cite{Hertzberg:2010yz}. The answer to this question, and indeed the preceding one, would be useful in searching for signatures of these objects in the early Universe, observations of which may act as smoking-guns for galileon inflation or galilean genesis.

Given the $C^1$ nature of the solutions, there is also the question of how ubiquitous these objects are. Do they form from from generic initial data or do they require special tunings to be realized? From the point of view of PDEs, these features are not new and it is often the case that shock fronts or $C^0$-peaked traveling waves form spontaneously in fluids and other non-linear systems. This question, is particularly pertinent because it also relates to the feasibility of forming such objects in the early Universe. Indeed, the formation of oscillons from the fragmentation of the homogeneous condensate can dominate the equation of state in the early Universe \cite{Amin:2010dc,Lozanov:2017hjm}.

Another important consideration is how these objects fit into the effective field theory of massive galileons. In order to find small-amplitude solutions analytically it was necessary to take $m\gg\Lambda$ ($g_3$, $g_4\gg1$), which is certainly peculiar from an EFT perspective. Indeed, a mass larger than the cut-off implies that new light states may be present that we have not accounted for. Similarly, the higher-order operators in the Wilsonian effective action (equation \eqref{eq:wilson}) scale as $(\partial/\Lambda)^p(\partial^2\pi/\Lambda^3)^q\sim (m/\Lambda)^{p+3q}$, and are therefore more relevant than the galileon operators, and so should have been included. This issue also arises when considering the Vainshtein mechanism, but in a slightly different context. In particular, operators with very large $p$ and $q$ become increasingly important near the Vainshtein radius \cite{Kaloper:2014vqa}. One possible resolution is that the galileons do not need additional UV-physics to preserve unitarity, so that the IR-theory does not contain any higher-derivative terms \cite{Keltner:2015xda}. Another is that the dimensionless coefficients of the higher-derivative terms are $\ll1$, contrary to what one would expect, due to a re-ordering of the counter-terms to preserve an approximate symmetry that is present in the limit where the galileons dominate over the canonical kinetic term \cite{Nicolis:2004qq,Kaloper:2014vqa}. Said another way, the coefficients are such that the operators resum into some simple object dictated by the approximate symmetry. Regardless, we view the condition $m\gg\Lambda$ not as being necessary for the existence of oscillons, but rather as a convenient limit that allows for their analytic construction in a systematic manner in order to demonstrate their existence, at least in this limit. As discussed a number of times above, this is analogous to the procedure used to construct small-amplitude flat-top oscillons analytically, and in that case one typically finds oscillons ubiquitously for the entire parameter space; the difference being that they must be found numerically. It would be interesting to solve the galileon equation numerically for this purpose, but such a study lies outside the scope of this work. 

Throughout this work we have restricted ourselves to galileons defined on Minkowski space, our motivation being to find breather solutions in the simplest possible setting. The study of oscillons on de Sitter space is particularly interesting because it leads to radiating tails \cite{Farhi:2007wj} and it would be interesting to investigate how higher-derivative oscillons behave on curved backgrounds for this reason. Unlike potential and $P(X)$-supported oscillons, galileons as defined on Minkowski space are not a covariant theory (the exception being the cubic) and the quartic galileon that was essential for supporting the oscillons we found in this work does not have a unique covariantization. One must choose from several such as covariant galileons \cite{Deffayet:2009wt}, beyond Horndeski galileons \cite{Gleyzes:2014dya,Gleyzes:2014qga}, massive gravity etc. We have not done so in this work since this would have required us to investigate a more complicated theory. Another approach would be to use different galileons defined on these spaces e.g. de Sitter galileons \cite{Burrage:2011bt} or more general theories \cite{Goon:2011qf}.

Finally, the connection of massive galileon EFTs to massive gravity raises the tantalizing prospect that such objects may exist within massive gravity theories themselves. In this preliminary investigation we have discovered these objects and analyzed their basic properties. A future investigation into their stability and formation dynamics could reveal a plethora of observational tests. Such investigations would likely be heavily numerical. 

\section*{Acknowledgements}

We are extrememly grateful for conversations with Mustafa Amin, Mark Hertzberg, and Austin Joyce. JS is supported by funds provided by the Center for Particle Cosmology. MT is supported in part by US Department of Energy (HEP) Award DE-SC0013528, and by NASA ATP grant NNH17ZDA001N.

\appendix

\section{Phase Space Analysis for the Cubic Galileon in 1 + 1 Dimensions}
\label{app:cubic}

\begin{figure}[ht]
\centering
\includegraphics[width=0.75\textwidth]{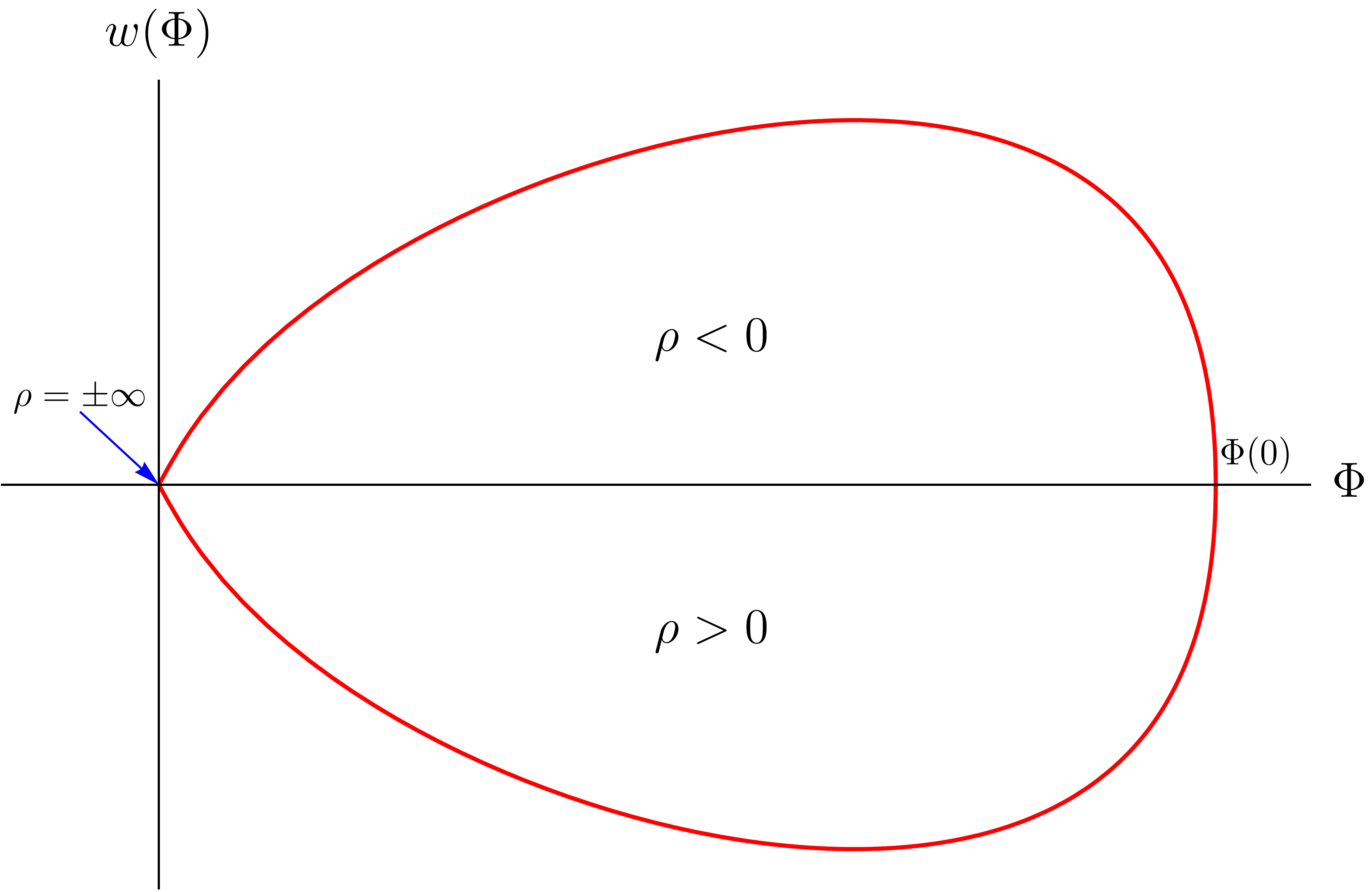}
\caption{An example of the phase space of equation \eqref{eq:weqncubic}. The solution at $\rho=\pm\infty$ is mapped onto the origin $w(\so)=w'(\so)=0$, which follows from the relation \eqref{eq:rholarge1d}. The same relation gives $w(\so)=-\so$ for $\rho\rightarrow\infty$, so beginning the integration at the origin (in $\so$--$w$ space) and taking this condition gives the solution for large positive $\rho$. In particular, since localized solutions have $\so'(\rho)<0$ for $\rho>0$ the curve moves into the lower half-plane where $w(\so)=\so'(\rho)<0$. As one moves closer to $\rho=0$ the red curve moves towards the axis. The point where it crosses i.e. $w=0$, $\so=\so(0)$ corresponds to $\rho=0$, since the boundary condition $\so'=0$ ($w=0$) is satisfied. The upper-half plane of phase space corresponds to $\rho<0$, since $w=\so'(\rho)>0$ for localized solutions and so at this point the red curve moves back towards the origin, arriving with a gradient $w'(\so)\sim+1$ corresponding to the $\rho\rightarrow-\infty$ solution in equation \eqref{eq:rholarge1d}.  }\label{fig:wex}
\end{figure}
 
In this appendix we provide the technical details of how localized solutions of equation \eqref{eq:firstintcubic} (with $\mathcal{E}=0$) can be constructed numerically. The strategy will be to work in the phase space $\{\so,\,\so'\}$ rather than the configuration space $\{\rho,\,\so(\rho)\}$. In particular, performing the change of variable:
\begin{equation}
\so'(\rho)=w(\so),\label{eq:w1} 
\end{equation}
the equation for the oscillon profile becomes
\begin{equation}\label{eq:weqncubic}
\frac{5}{6}\so^2w^3\frac{\dd w}{\dd\Phi_1^2}+\frac{5}{12}\so^2w^2\left(\frac{\dd w}{\dd\so^2}\right)^2+\frac{5}{3}\so w^3\frac{\dd w}{\dd{\so}}+w^2-\frac{w^4}{12}-\so^2+\frac{3\so^4}{8}=0.
\end{equation}
Note that we are now treating $w$ as a function of $\so$. (In this section we will use primes to denote derivatives of functions with respect to their arguments.) Not only have we reduced the equation from third to second-order, but, as we will now show, we have mapped the domain $\rho\in[-\infty,\infty]$ to $\so\in[0,\so(0)]$. This makes finding the profile far easier because the boundary condition at $\rho=\infty$ is now mapped onto the origin. In particular, as $\rho\rightarrow\infty$ we have equation \eqref{eq:rholarge1d} i.e. $\so\rightarrow0$ implying that the origin $\so=0$ corresponds to the points $\rho=\pm\infty$. Furthermore, differentiating equation \eqref{eq:rholarge1d} near $\rho=\pm\infty$ we have $w=\pm\so$ near the origin. This implies that the boundary conditions for equation \eqref{eq:weqncubic} are $w(\so=0)=0$ and $w'(\so=0)=-1$. The boundary conditions at $\rho=0$ now imply that an oscillon profile exists if one can integrate equation \eqref{eq:weqncubic} from the origin ($\so=0$) given these boundary conditions and find a second point where $w(\so)=0$. This point corresponds to the amplitude $\so(0)$ because here we have $w(\so)=0\Rightarrow\so'(\rho)=0$ with $\so\ne0$. The situation is exemplified in figure \ref{fig:wex}.

Setting $w=0$ in equation \eqref{eq:weqncubic} one finds two zeros: $\so=0$ corresponding to the origin and $\so(0)^2=\frac{8}{3\xi}$, which corresponds to $\so''(0)=0$ so that the profile is flat. Since neither of these corresponds to a physically acceptable solution we conclude that oscillon solutions only exist if the curve $w(\so)$ crosses the $\so$ axis at some value $\so(0)<\sqrt{{8}/{3}}$. Whether or not this is the case requires a numerical solution.\footnote{One can try to take limits of equation \eqref{eq:weqncubic} where combinations of $w'(\so)$ and its derivatives are fixed to satisfy \eqref{eq:origincubic} but only the trivial solutions can be found. Intuitively, this is because $\so(0)$ is determined by the requirement that the solution obeys the boundary conditions at $\so=0$ ($\rho\rightarrow\infty$) so its value is dependent on both the boundary conditions and the differential equation so that it cannot be determined from the latter alone.} Furthermore, $\so''(\rho)$ is finite and non-zero by virtue of equation \eqref{eq:origincubic}, and taking derivatives of equation \eqref{eq:w1} implies that the product $w'(\so(0))w(\so(0))$ is finite. (The condition $\so^{(3)}(0)=0$ also implies $w''(\so)=-w'({\so})^2/w$.) Since $w=0$ at this point the curve must approach it with $w'(\so)\rightarrow\infty$. The oscillon profile can be found by integrating equation \eqref{eq:w1}.

\bibliography{ref}

\end{document}